\documentclass[conference]{IEEEtran}
\usepackage[left=0.635in,right=0.635in,top=0.75in,bottom=1.05in]{geometry}

\usepackage[numbers]{natbib}
\usepackage{caption}
\usepackage{subcaption}
\usepackage{multirow}
\usepackage{tabularray}
\usepackage{algorithmic}
\usepackage{graphicx}
\usepackage{textcomp}
\usepackage{fancyhdr}
\usepackage[table]{xcolor} 
\usepackage{colortbl}
\usepackage{makecell}
\usepackage{tabularx}
\usepackage{float}
\usepackage{amsmath}
\usepackage{amssymb}
\usepackage{amsfonts}
\usepackage{listings}
\usepackage{xcolor}
\usepackage{booktabs}
\usepackage{pifont}
\usepackage[most]{tcolorbox}
\usepackage{subcaption}
\usepackage{tikz}
\usepackage{caption}
\usepackage{enumitem}
\usepackage{url}
\usepackage{subcaption} 
\usepackage{booktabs}   
\usepackage{multirow}   
\usepackage{stfloats}

\definecolor{PineGreenDarkest}{HTML}{4C9E8F}
\definecolor{PineGreenDark}{HTML}{70B3A1}
\definecolor{PineGreenMedium}{HTML}{94C8B4}
\definecolor{PineGreenLight}{HTML}{B8DDC6}
\definecolor{PineGreenLighter}{HTML}{D1EED9}
\definecolor{darkgreen}{rgb}{0.0, 0.5, 0.0} 
\definecolor{mygreen}{HTML}{afcb0c}

\lstdefinestyle{mypython}{
    language=Python,
    basicstyle=\ttfamily\small,
    keywordstyle=\color{blue}\bfseries,
    stringstyle=\color{darkgreen},
    commentstyle=\color{gray}\itshape,
    morekeywords={BaseModel, Field, List},
    columns=flexible,
    keepspaces=true,
    breaklines=true,
    showstringspaces=false,
    frame=single,
    rulecolor=\color{black},
    stepnumber=1,
    numbersep=5pt,
    xleftmargin=0.05\columnwidth,   
    xrightmargin=0.05\columnwidth,  
}

\begin{document}
%
\title{SIMCODE: A Benchmark for  Natural Language to {\color{mygreen}\textbf{ns-3}} Network Simulation Code Generation}

\author{
  \IEEEauthorblockN{Tasnim Ahmed, Mirza Mohammad Azwad, Salimur Choudhury}
  \IEEEauthorblockA{School of Computing, Queen's University, Ontario, Canada\\
    \{tasnim.ahmed, mizra.mohammadazwad, s.choudhury\}@queensu.ca}
}

\IEEEspecialpapernotice{This paper has been accepted for presentation at The $50^{th}$ IEEE Conference on Local Computer Networks (LCN). Special Track on Large Language Models and Networking}

\maketitle

\begin{abstract}

Large language models (LLMs) have demonstrated remarkable capabilities in code generation across various domains. However, their effectiveness in generating simulation scripts for domain-specific environments like ns-3 remains underexplored. Despite the growing interest in automating network simulations, existing tools primarily focus on interactive automation over rigorous evaluation. To facilitate systematic evaluation, we introduce SIMCODE, the first benchmark to evaluate LLMs' ability to generate ns-3 simulation code from natural language. SIMCODE includes 400 tasks across introductory, intermediate, and advanced levels, with solutions and test cases. Using SIMCODE, we evaluate three prominent LLMs, Gemini-2.0, GPT-4.1, and Qwen-3, across six prompt techniques. Furthermore, investigating task-specific fine-tuning’s impact reveals that while GPT-4.1 outperforms others, execution accuracy remains modest, with substantial room for improvement. Error analysis identifies missing headers and API mismatches as dominant failures. Nevertheless, SIMCODE provides a foundational step toward evaluating LLMs and research in domain-aware generative systems.

\end{abstract}

\IEEEpeerreviewmaketitle

\section{Introduction}
Large Language Models (LLMs) have revolutionized code generation, facilitating developers and researchers to translate natural language prompts into executable code across various programming languages and domains. From general-purpose coding benchmarks, such as HumanEval \cite{chen2021evaluating} and MBPP \cite{austin2021programsynthesislargelanguage}, to domain-specific evaluations, including MojoBench \cite{raihan2024mojobench} and mHumanEval \cite{raihan2024mhumaneval}, the field has witnessed a surge in interest in assessing LLMs' capabilities in real-world programming tasks. Despite these advances, a significant gap remains in evaluating LLMs in simulation-driven, domain-specific environments such as network simulation. Simulating network scenarios requires syntactically correct code, domain knowledge, protocol awareness, and runtime correctness.

ns-3 \cite{riley2010ns}, a widely adopted discrete-event network simulator, is central to academic and industrial research in wired and wireless networking. However, scripting simulations in ns-3 involve steep learning curves due to its use of C++, adherence to telecom standards like 3GPP, and complex protocol stacks. Recent works such as GenOnet \cite{genonet2024} and LangChain-based simulation frameworks \cite{rezazadeh2025toward} have begun integrating LLMs with ns-3, offering automated tools to generate and execute simulation scripts using natural language. These tools demonstrate the feasibility of using LLMs to simulate 5G/6G networks, yet they primarily focus on interactive automation rather than systematic evaluation.

\begin{figure*}
    \centering
    \includegraphics[width=0.93\linewidth]{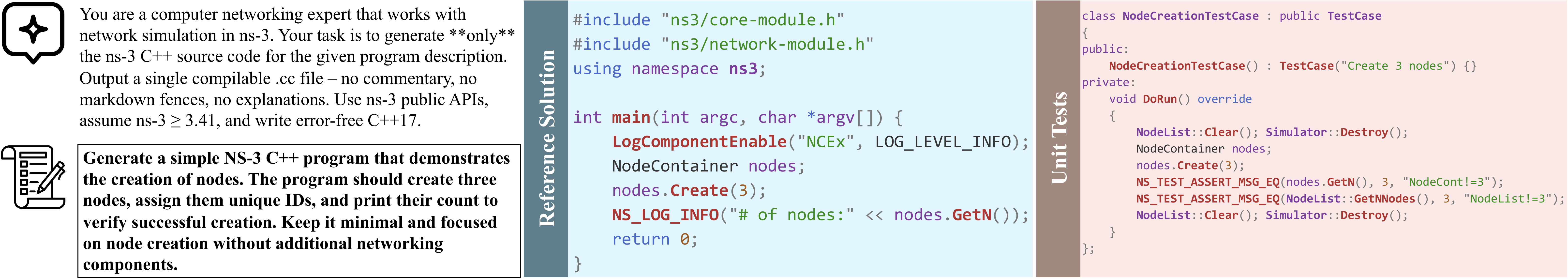}
    \caption{Structure of a SIMCODE benchmark sample, comprising a natural language prompt, ns-3 C++ solution, and test cases.
}
    \label{fig:simcode}
\end{figure*}

To bridge this research gap, we introduce SIMCODE, a structured benchmark designed to evaluate the ability of LLMs to generate correct and executable ns-3 simulation code from natural language prompts. Unlike tool-assisted frameworks, SIMCODE provides a static, curated dataset grounded in real-world networking scenarios and educational courses.
Each sample in SIMCODE (as shown in Figure \ref{fig:simcode}) is represented as a triplet: a natural language prompt, a verified C++ solution, and a set of test cases to validate simulation correctness. The contributions of this study are as follows.

\begin{itemize}
    \item SIMCODE is introduced as the first benchmark to evaluate LLMs' ability to generate executable ns-3 simulation code from natural language.

    \item It includes $400$ tasks across introductory, intermediate, and advanced levels, each with a prompt, verified C++ solution, and test cases.

    \item Evaluation of Gemini-2.0, GPT-4.1, and Qwen-3 across six prompt techniques shows GPT-4.1 achieving up to 30.6\% execution accuracy, with task-specific fine-tuning enhancing performance.

    \item Error analysis reveals issues like missing headers and API mismatches, recommending compilation feedback and retrieval-based systems to improve code generation.
\end{itemize}

The remainder of this paper is organized as follows. Section \ref{sec:related} reviews the related work. Section \ref{sec:dataset} describes the SIMCODE benchmark and its construction. Section \ref{sec:exp} presents the experimental setup, model selection, prompt techniques, and evaluation metrics. Section \ref{sec:res} discusses the results and insights. Section \ref{sec:error} analyzes the most frequent errors and outlines limitations and future directions. Section \ref{sec:conclusion} concludes with a summary of findings and their implications for advancing LLM-driven network simulation.

\section{Related Works}
\label{sec:related}
LLMs have rapidly advanced programming assistance by facilitating functional code generation from natural language. A landmark achievement was OpenAI Codex \cite{manning2022researchagenda}, fine‑tuned on GitHub code, which powers GitHub Copilot and achieved a $28.8\%$ solve rate on the HumanEval benchmark of $164$ Python problems \cite{chen2021evaluating}. Shortly thereafter, the MBPP dataset (Mostly Basic Programming Problems) provided $974$ crowd‑sourced Python tasks, demonstrating that few‑shot prompting of LLMs could solve over $50\%$ of these problems \cite{austin2021programsynthesislargelanguage}. Subsequent benchmarks such as APPS and CodeNet further broadened evaluation across diverse problem sets and programming languages \cite{codenet2021}. These foundational efforts established standardized metrics and baselines for assessing code‑generation quality.

Building upon early benchmarks, researchers introduced feedback‑driven and agentic workflows to improve reliability. Fakhoury et al. \cite{fakhoury2024} proposed the TiCoder interactive framework, where users iteratively validate generated tests to refine code, achieving a $46\%$ gain in pass@1 accuracy after five rounds. Huang et al. \cite{huang2024agentcodermultiagentbasedcodegeneration} presented AgentCoder, a multi‑agent system comprising programmer, test‑designer, and test‑executor agents; by generating and executing tests in a closed loop, AgentCoder achieved $77–89\%$ pass@1 accuracy on HumanEval variants.
Self-Refine \cite{madaan2023selfrefineiterativerefinementselffeedback}, for instance, shows that letting an LLM critique and re-write its own draft can improve pass@1 scores by double-digit margins without human intervention.
Another direction focuses on domain‑specific feedback. Kumar et al. \cite{kumar2025tfhecoderevaluatingllmagenticfully} introduced TFHE‑Coder for Fully Homomorphic Encryption code, adopting compiler‑in‑the‑loop and retrieval‑augmented prompts to reduce errors and improve code fidelity. These approaches (interactive test‑driven systems, multi‑agent loops, and compiler‑augmented pipelines) demonstrate that integrating runtime verification into LLM workflows significantly enhances output correctness. Whereas the above work remains focused in conventional software libraries, simulation frameworks such as ns-3 expose event-driven APIs and telecom-standard data models, making code synthesis uniquely challenging. Recent efforts therefore integrate LLMs directly with network simulators. Rezazadeh et al. \cite{genonet2024} proposed GenOnet, a multi‑agent architecture that integrates GPT models with 3GPP and O‑RAN standards to generate, execute, and analyze 5G/6G scenarios in ns‑3. Extending this, Mavroudis et al. \cite{langchain2024} proposed a LangChain‑based pipeline with specialized agents for script generation, test design, execution, and result interpretation. In their case study on the ns‑3 5G‑LENA module, the system converged to correct simulation code in two iterations on average, maintaining low syntax error rates. These demonstrations highlight the promise of LLM-assisted simulation yet each evaluates a single pipeline, so we still lack a unified benchmark for rigorous comparison. To this end, our research benchmarks multiple state-of-the-art LLMs across diverse ns-3 tasks. We aim to quantify modern LLMs’ capabilities in automating network simulation code by evaluating functional accuracy, generalization, and usability.

\section{The SIMCODE Benchmark}
\label{sec:dataset}
\subsection{Dataset Construction}
Our dataset construction consists of three key stages: (i) Simulation Task Collection, (ii) Prompt, Solution, and Test Case Creation, and (iii) Quality Assurance and Review. We detail each stage below:

\subsubsection{Simulation Task Collection}
The SIMCODE benchmark prioritizes diversity in both networking concepts and difficulty levels. We curated simulation tasks from various sources, including university-level networking coursework, the official ns-3 example repository\footnote{\url{https://www.nsnam.org/docs/release/3.45/tutorial/ns-3-tutorial.pdf}}, and original scenarios developed by the authors. These tasks cover a wide range of networking scenarios, including routing protocols, network topologies, and wireless networks. The tasks were grouped into Introductory, Intermediate, and Advanced levels according to their complexity and the length of their solutions, in alignment with standard coding benchmarks \cite{hendrycks2021measuring}. Each level is subdivided into subtopics, each subtopic originating from a core concept in network simulation. Introductory tasks focus on fundamental scenarios, such as establishing point-to-point links, creating nodes and simple topologies, and implementing basic communication protocols. Intermediate tasks involve complex configurations, including multi-hop topologies with traffic flows, advanced point-to-point links, large-scale Wi-Fi and LTE networks, and more intricate protocol examples. Advanced tasks require a high level of expertise and cover activities such as developing custom protocols, optimizing network performance under constraints, implementing congestion control mechanisms, configuring CSMA, simulating Mobile Ad Hoc Networks, managing socket connections, and designing matrix topologies. In total, the dataset includes $400$ simulation tasks. The distribution of tasks across difficulty levels is presented in Table~\ref{tab:dataset-by-category}, and the distribution by subtopics is detailed in Table~\ref{tab:cot_hier}.

\subsubsection{Prompt, Solution, and Test Case Creation}
For each simulation task, we created a triplet consisting of a natural language prompt, a reference code solution, and a set of test cases, adhering to the prompt–code–test structure of existing benchmarks \cite{hendrycks2021measuring, austin2021programsynthesislargelanguage}. All co-authors with expertise in network simulation and competitive programming collaboratively developed these components.\\
\textbf{Prompts}: Clear, unambiguous natural language descriptions of the simulation tasks were constructed, averaging approximately $250$ tokens.\\
\textbf{Solutions:} Reference solutions were implemented in C++ using ns-3. Each solution was verified to run correctly in the ns-3 environment and follows best practices for simulation scripting. Solution lengths vary by difficulty, averaging $40$ lines for introductory tasks, $80$ for intermediate tasks, and over $100$ for advanced tasks.\\
\textbf{Test Cases:} Each problem includes a rigorous set of unit test cases, averaging $5–7$ per task designed to validate simulation outputs. Implemented using ns-3’s tracing mechanisms, these tests assert conditions such as node count, packet delivery ratios, throughput thresholds, etc. They cover both normal behaviour and edge cases, ensuring comprehensive evaluation.

\subsubsection{Quality Assurance and Review}
A thorough quality assessment was conducted to ensure the accuracy and reliability of the dataset. Each data point was independently reviewed by at least two authors, who collectively bring academic experience from undergraduate and graduate-level networking and advanced networking courses. Prompts were evaluated for clarity and completeness, ensuring all necessary details were provided without ambiguity. Solutions went through code reviews to confirm correctness, efficiency, and adherence to ns-3 standards, with execution tests verifying simulation outcomes. Test Cases were validated to ensure they accurately assessed both correct implementations and standard errors, using reference solutions and erroneous variants.

\subsection{Dataset Analysis}
\subsubsection{Task Diversity}
The SIMCODE dataset comprises a broad range of networking simulation tasks that reflect diversity in conceptual coverage and technical complexity. As shown in Table 1, tasks are distributed across difficulty levels, with varying solution lengths and test case counts.

\begin{table}[h]
\centering
\caption{Summary of Dataset Characteristics}
\begin{tabular}{l c c c}
\toprule
\textbf{Category (Difficulty)} & \textbf{\# Problems} & \textbf{Avg. Sol. Length} & \textbf{Avg. \# Tests} \\
\midrule
Introductory & 111 & $\sim$60 & 3--5\\
Intermediate & 101 & $\sim$110 & 5--8\\
Advanced & 189 & 150+ & 8--10\\
\bottomrule
\end{tabular}
\label{tab:dataset-by-category}
\end{table}

\subsubsection{Linguistic and Conceptual Diversity}
The natural language prompts exhibit significant linguistic diversity, mirroring real-world task descriptions. With an average length of $250$ tokens, they provide detailed specifications comparable to APPS \cite{hendrycks2021measuring}, which averages $293$ words per prompt. A type-token ratio (TTR) analysis of the dataset provides a value of approximately $0.69$, which reflects a considerable degree of lexical diversity. The dataset covers various networking tasks, from basic connectivity scenarios to complex protocol-level design simulations. Such variety necessitates programming proficiency and a strong foundation in domain-specific networking knowledge. Unlike general-purpose coding benchmarks, SIMCODE is a unique and challenging resource for evaluating code generation in specialized domains.

\section{Experiments}
\label{sec:exp}
\subsection{Model Selection}


We experiment with SIMCODE using three models, comprising a mix of proprietary and open-source models for code generation to gather broader insights. Additionally, we include one fine-tuned model in our evaluation to assess the impact of task-specific adaptation. The proprietary models include Gemini-2.0-flash and GPT-4.1, while Qwen3 is open-source. We also evaluate a fine-tuned variant of GPT-4.1, adapted using SIMCODE training samples.

\subsection{Prompt Engineering}
Prompt design is a primary lever for steering LLMs toward faithful, compilable code.  Inspired by prior work on controlled generation \cite{10.1145/3690635, 10.1145/3672456}, we evaluate six template families that progressively enrich the input with structural cues, exemplars, or self-verification signals. The prompt templates are summarized in Table~\ref{tab:prompt-templates}. All prompts (i) cast the model as an ns-3 specialist, (ii) forbid commentary or markdown, and (iii) mandate C++17 compatibility for ns-3 ver. $\geq3.41$.  The templates differ, however, in the auxiliary scaffolding they provide.

\begin{table*}[htbp]
  \centering
  \caption{Taxonomy of prompt designs and their intended improvements in code generations}
  \label{tab:prompt-templates}
  \begin{tabularx}{\textwidth}{@{}lX@{}}
    \toprule
    \textbf{Prompt template} & \textbf{Description} \\
    \midrule
    Instruction & A bare prompt supplies only the minimal directive
                  “generate \textit{ns-3} C++ code.”  
                  It serves as a lower-bound baseline, absent any intermediate
                  reasoning or examples. \\[0.4em]

    Chain-of-Thought (CoT) \cite{10.5555/3600270.3602070} &
      Encourages the LLM to reason step-by-step about simulation components
      (nodes, network devices, protocols, parameters) before producing code,
      aiming for greater logical coherence and accuracy. \\[0.4em]

    Few-Shot &
      Provides a solved example that maps a simulation description to its code,
      helping the LLM generalize.  
      In our preliminary runs, it consistently improves header selection and
      simulation cleanup routines. \\[0.4em]

    ReAct \cite{yao2023reactsynergizingreasoningacting} &
      Employs a two-phase strategy: first reasoning about topology,
      configurations, and protocols, then “acting” by emitting the 
      corresponding code. \\[0.4em]

    Expert Prompt &
      Primes the model as a “renowned \textit{ns-3} developer,” explicitly
      requesting best practices (precise attribute configuration, helper
      usage).  
      By invoking domain-specific proficiency, it targets professional-grade
      simulation code. \\[0.4em]

    Self-Consistency \cite{wang2023selfconsistencyimproveschainthought} &
      Embeds a draft–critique–rewrite loop (an enhanced CoT).  
      The model drafts, audits for common \textit{ns-3} pitfalls (missing
      headers, deprecated APIs, logical flaws), and outputs a corrected
      version. \\
    \bottomrule
  \end{tabularx}
\end{table*}

\subsection{Code Execution}
After generating the code, we move to execution. LLMs produce well-structured code blocks that require minimal cleanup. We use simple RegEx commands to extract and evaluate these blocks locally in a WSL environment that runs ns-3.

\subsection{Evaluation Metrics}
A comprehensive set of metrics has been adopted to evaluate the functional correctness and the structural similarity of the generated code concerning the reference solutions. Execution accuracy is paramount, as it is a binary metric that measures the proportion of generated code samples that successfully compile and execute. Complementing this, CodeBLEU \cite{ren2020codebleumethodautomaticevaluation} incorporates n-gram matching, weighted n-gram matching, abstract syntax tree matching, and data-flow analysis. These components collectively assess the syntactic structure and semantic correctness of the generated code. Additionally, CodeBERTScore utilizes embeddings derived from the CodeBERT\footnote{\url{https://huggingface.co/microsoft/codebert-base}} model to compute the semantic similarity between the generated and reference code, thereby capturing deeper contextual and functional equivalences that extend beyond surface-level syntax.
Finally, the Pass@1 is another binary metric that quantifies the percentage of instances in which the first generated code sample is correct and passes all the respective unit tests.

\section{Results \& Insights}
\label{sec:res}
Table \ref{tab:simcode-main} presents the performance of each model on the SIMCODE dataset across six distinct prompt strategies. Results show that GPT‐4.1 is the best-performing baseline model on \textsc{SIMCODE}.  Its CoT prompt reaches $29.3\%$ execution accuracy, and its average across all six prompts is $27.7\%$.  Qwen-3 stands in the middle with an average of $24.3\%$, and Gemini stays near $15\%$.  The gaps in CodeBLEU and CodeBERTScore are much smaller because these two metrics reward token overlap and correct header usage, which all models manage to produce.  When we move from these static measures to real execution, the difference widens, showing that correctness at run time needs stronger reasoning than the open-source models presently offer. Furthermore, Pass@1 is a tighter metric than execution accuracy, resulting in a substantial drop in LLM performance. However, upon analyzing the generated codes, we found that test cases often fail not because the solutions are incorrect, but because the implementation logic deviates from the reference, even when functionally correct. These findings lead to three practical guidelines for future users. First, the choice of model has a greater impact on execution accuracy than the prompt design. GPT-4.1 consistently outperforms Gemini-2.0 and Qwen-3, even when using a basic instruction prompt. Second, CoT is the most consistent prompting strategy across models, though its improvements are relatively small for strong models and vary across weaker ones. This indicates that adjusting the prompt alone may not lead to significant improvements. Third, even a lightweight domain-specific fine-tuning brings noticeable improvements and can be viable when dependable code generation is needed.

{
\setlength{\tabcolsep}{3pt}
\begin{table}[h]
\centering
\scriptsize
\caption{Performance of LLMs on each prompt type.}
\label{tab:simcode-main}
\begin{tabular}{llcccc}
\toprule
\textbf{Model} & \textbf{Prompt} & \textbf{CodeBLEU} & \textbf{CodeBERT} & \textbf{Exec.\ Acc.} & \textbf{Pass@1} \\
\midrule
\multirow{6}{*}{Gemini-2.0}
 & Instruction            & 0.732 & 0.927 & 0.133 & 0.055 \\
 & CoT                    & 0.731 & 0.927 & 0.155 & 0.055 \\
 & Few-shot               & 0.722 & 0.932 & 0.190 & 0.065 \\
 & ReAct                  & 0.726 & 0.928 & 0.143 & 0.045 \\
 & Expert                 & 0.720 & 0.926 & 0.205 & 0.115 \\
 & Self-consistency       & 0.743 & 0.929 & 0.065 & 0.010  \\
\midrule
\multirow{6}{*}{GPT-4.1}
 & Instruction            & 0.760 & 0.930 & 0.265 & 0.088 \\
 & CoT                    & 0.763 & 0.931 & 0.293 & 0.153 \\
 & Few-shot               & 0.765 & 0.934 & 0.280 & 0.140 \\
 & ReAct                  & 0.765 & 0.931 & 0.283 & 0.140 \\
 & Expert                 & 0.756 & 0.931 & 0.258 & 0.123 \\
 & Self-consistency       & 0.773 & 0.929 & 0.285 & 0.133 \\
\midrule
\multirow{6}{*}{Qwen-3}
 & Instruction            & 0.733 & 0.934 & 0.260 & 0.163 \\
 & CoT                    & 0.732 & 0.933 & 0.263 & 0.193 \\
 & Few-shot               & 0.732 & 0.935 & 0.220 & 0.100 \\
 & ReAct                  & 0.724 & 0.933 & 0.245 & 0.125 \\
 & Expert                 & 0.729 & 0.933 & 0.248 & 0.122 \\
 & Self-consistency       & 0.748 & 0.933 & 0.223 & 0.095 \\
 \midrule
\multirow{1}{*}{\textbf{GPT-4.1 (FT)}} & \textbf{Instruction} & \textbf{0.785} & \textbf{0.940} & \textbf{0.306} & -- \\
\bottomrule
\end{tabular}
\end{table}
}


\subsection{Prompt sensitivity}
\begin{figure}
    \centering
    \includegraphics[width=\linewidth]{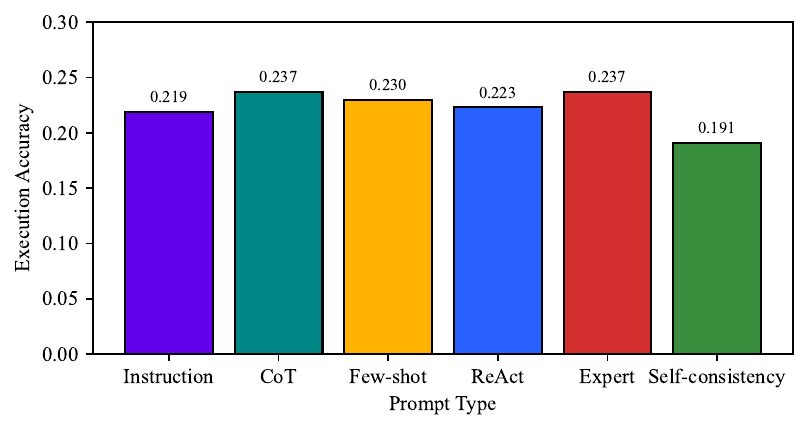}
    \caption{Average execution accuracy per prompt type.}
    \label{fig:prompt_acc}
\end{figure}

Prompt choice changes execution accuracy more than any similarity score in Table \ref{tab:simcode-main}.
For every LLM, CoT is better than the plain instruction prompt, giving about $2$ to $3$ percentage extra accuracy.  Few-shot examples help Gemini the most, adding around five percentage points because concrete demonstrations teach the model how to use \textit{ns-3} helper classes.  Self-consistency performs well for GPT-4.1, yet it hurts Gemini and Qwen-3. An intriguing finding from our study is that Qwen-3 demonstrates strong performance when provided with a basic instruction prompt, and applying additional prompt engineering techniques does not yield substantial improvements. Figure \ref{fig:prompt_acc} illustrates the average execution accuracy for each prompt type, aggregated across all baseline LLMs.

\subsection{Similarity metrics}
The CodeBLEU and CodeBERTScore values in Table \ref{tab:simcode-main} remain close across all models and prompts.  Both metrics look mainly at surface overlap and shared abstract syntax-tree fragments.  Almost every generated file includes the same block of \verb|#include| statements and the same pattern of node creation.  Since these parts dominate the embeddings, CodeBERTScore remains relatively constant, ranging between $0.926$ and $0.94$, regardless of the model used. CodeBLEU shows a marginally higher sensitivity to differences across models.
Thus, similarity metrics give an illusion of uniform quality, while execution accuracy exposes the real functional gap.

\subsection{Performance Across Task Complexity and Network Domains}
\begin{table*}[t]
\centering
\caption{Execution accuracy by difficulty and topic under the CoT prompt for all baseline models}
\label{tab:cot_hier}
\small
\setlength{\tabcolsep}{4pt}
\renewcommand{\arraystretch}{1.05}
\begin{tabular}{l l c c c c c c c c c c}
\toprule
\textbf{Difficulty} & \textbf{Topic} & \textbf{Total} & \multicolumn{3}{c}{\textbf{Gemini-2.0}} & \multicolumn{3}{c}{\textbf{GPT-4.1}} & \multicolumn{3}{c}{\textbf{Qwen-3}} \\
\cmidrule(lr){4-6}
\cmidrule(lr){7-9}
\cmidrule(lr){10-12}
& & & \textbf{Correct} & \textbf{Accuracy} & \textbf{Average} & \textbf{Correct} & \textbf{Accuracy} & \textbf{Average} & \textbf{Correct} & \textbf{Accuracy} & \textbf{Average} \\\midrule
\multirow{17}*{\textbf{Advanced}} & Congestion & 9 & -- & -- & \multirow{17}*{0.143} & -- & -- & \multirow{17}*{0.259} & 1 & 0.111 & \multirow{17}*{0.233} \\
 & CSMA & 11 & 3 & 0.273 &  & 4 & 0.364 &  & 1 & 0.091 &  \\
 & IP & 14 & 2 & 0.143 &  & 5 & 0.357 &  & 4 & 0.286 &  \\
 & LAN & 12 & -- & -- &  & 1 & 0.083 &  & 1 & 0.083 &  \\
 & LTE & 2 & -- & -- &  & 2 & 1.000 &  & 1 & 0.500 &  \\
 & Manet & 17 & -- & -- &  & -- & -- &  & -- & -- &  \\
 & Matrix & 1 & -- & -- &  & 1 & 1.000 &  & -- & -- &  \\
 & Nodes & 5 & 2 & 0.400 &  & 3 & 0.600 &  & 2 & 0.400 &  \\
 & P2P & 14 & 6 & 0.429 &  & 11 & 0.786 &  & 11 & 0.786 &  \\
 & Protocols & 15 & -- & -- &  & -- & -- &  & -- & -- &  \\
 & Socket & 2 & -- & -- &  & 2 & 1.000 &  & 1 & 0.500 &  \\
 & TCP & 17 & 4 & 0.235 &  & 8 & 0.471 &  & 9 & 0.529 &  \\
 & Topology & 23 & 7 & 0.304 &  & 9 & 0.391 &  & 9 & 0.391 &  \\
 & UDP & 5 & 3 & 0.600 &  & 2 & 0.400 &  & 4 & 0.800 &  \\
 & Vanet & 7 & -- & -- &  & -- & -- &  & -- & -- &  \\
 & WAN & 2 & -- & -- &  & 1 & 0.500 &  & -- & -- &  \\
 & Wifi & 33 & -- & -- &  & -- & -- &  & -- & -- &  \\
\midrule
\multirow{5}*{\textbf{Intermediate}} & LTE & 9 & -- & -- & \multirow{5}*{0.109} & 2 & 0.222 & \multirow{5}*{0.277} & -- & -- & \multirow{5}*{0.257} \\
 & P2P & 14 & 7 & 0.500 &  & 13 & 0.929 &  & 10 & 0.714 &  \\
 & TCP & 14 & 2 & 0.143 &  & 11 & 0.786 &  & 4 & 0.286 &  \\
 & UDP & 11 & 2 & 0.182 &  & 2 & 0.182 &  & 1 & 0.091 &  \\
 & Wifi & 53 & -- & -- &  & -- & -- &  & 11 & 0.208 &  \\
\midrule
\multirow{14}*{\textbf{Introductory}} & CSMA & 8 & 6 & 0.750 & \multirow{14}*{0.216} & 7 & 0.875 & \multirow{14}*{0.360} & 5 & 0.625 & \multirow{14}*{0.315} \\
 & IP & 3 & -- & -- &  & -- & -- &  & -- & -- &  \\
 & LAN & 2 & 1 & 0.500 &  & 1 & 0.500 &  & -- & -- &  \\
 & LTE & 8 & -- & -- &  & 1 & 0.125 &  & -- & -- &  \\
 & Manet & 6 & -- & -- &  & -- & -- &  & -- & -- &  \\
 & Nodes & 12 & 3 & 0.250 &  & 5 & 0.417 &  & 5 & 0.417 &  \\
 & P2P & 10 & 6 & 0.600 &  & 10 & 1.000 &  & 9 & 0.900 &  \\
 & Protocols & 7 & -- & -- &  & -- & -- &  & 1 & 0.143 &  \\
 & TCP & 11 & 3 & 0.273 &  & 8 & 0.727 &  & 2 & 0.182 &  \\
 & Topology & 11 & 5 & 0.455 &  & 6 & 0.545 &  & 1 & 0.091 &  \\
 & UDP & 9 & -- & -- &  & 2 & 0.222 &  & 2 & 0.222 &  \\
 & Vanet & 3 & -- & -- &  & -- & -- &  & -- & -- &  \\
 & WAN & 2 & -- & -- &  & -- & -- &  & -- & -- &  \\
 & Wifi & 19 & -- & -- &  & -- & -- &  & 10 & 0.526 &  \\
\midrule
\textbf{Overall} & -- & 400 & 62 & 0.155 & -- & 117 & 0.292 & -- & 105 & 0.263 & -- \\
\bottomrule
\end{tabular}
\vspace{-1em}
\end{table*}

Table \ref{tab:cot_hier} presents a breakdown of performance for the CoT (best performing prompt type) prompt across difficulty levels and network-related topics. The results highlight where each model performs well and where their limitations are most apparent. Across all difficulty levels, GPT-4.1 consistently outperforms the other models. The strongest results of GPT-4.1 come from topics such as LTE, Socket programming, and P2P communication. Gemini struggles significantly in almost every difficulty level, with many topics showing zero correct completions.
For advanced-level questions, GPT-4.1 achieves an overall accuracy of $25.9\%$, compared to $23.3\%$ for Qwen-3 and $14.3\%$ for Gemini.
All models show improved performance in the intermediate-level tasks, though the gap remains. GPT-4.1 achieves $27.7\%$ accuracy, followed by Qwen-3 at $25.7\%$ and Gemini at just $10.9\%$. For introductory-level tasks, the accuracy increases further, as expected. GPT-4.1 reaches $36.0\%$ accuracy, Qwen-3 follows with $31.5\%$, and Gemini improves to $21.6\%$. Despite these improvements, several topics remain entirely unsolved by every LLM.

\subsection{Impact of Fine-tuning}

To evaluate the effect of task-specific adaptation, we fine-tuned GPT-4.1. The dataset was partitioned into $60\%$ for training, $10\%$ for validation, and $30\%$ for testing. The fine-tuning was conducted over $2$ epochs with a batch size of $1$ and a learning rate multiplier of $2$. The model achieved a training loss of $0.180$ and a validation loss of $0.149$ without signs of overfitting. For evaluation, we used only the instruction prompt to isolate the impact of fine-tuning from prompt engineering. Table \ref{tab:simcode-main} reports the performance of the fine-tuned model. Fine-tuning achieves noticeable improvements across all evaluated metrics. Execution accuracy increases from $29.3\%$ (best prompt for GPT-4.1) to $30.6\%$, CodeBLEU rises from $0.773$ to $0.785$, and CodeBERTScore improves from $0.934$ to $0.940$. Results represent the average performance across $5$ separate runs. Although the relative gains may appear modest, they are consistent across all evaluations and demonstrate the value of fine-tuning for domain-specific code generation tasks.

\section{Analysis \& Future Directions}
\label{sec:error}

\subsection{Error Analysis}
\begin{table*}[!t]
  \centering
  \caption{Failure modes and recommended mitigations (Baseline GPT-4.1)}
  \label{tab:failure}
  \renewcommand{\arraystretch}{1.1}
  \begin{tabular}{@{}l r p{0.46\linewidth}@{}}
    \toprule
    \textbf{Failure mode} & \textbf{Share} & \textbf{Primary mitigation steps} \\ \midrule
    Missing or outdated header/module & 48\% &
      (i) State required modules or headers in the prompt; (ii) compile–feedback loop; (iii) retrieval of up-to-date header corpus. \\[2pt]
    API–mismatch / renamed method       & 21\% &
      (i) Compile–feedback loop; (ii) context-based retrieval with current APIs; (iii) targeted fine-tuning on recent ns-3 code. \\[2pt]
    Incorrect type / pointer            & 10\% &
      (i) Compile–feedback loop for type diagnostics; (ii) fine-tuning on type-safe examples. \\[2pt]
    Accessing private API               & 7\% &
      (i) Compile–feedback loop; (ii) fine-tuning to reinforce public-API usage. \\[2pt]
    Runtime failure                     & 5\% &
      (i) Compile–feedback loop extended to runtime tests; (ii) fine-tuning with execution-verified samples. \\[2pt]
    Miscellaneous (unused vars, CMake config errors, etc.) & 9\% &
      (i) Compile–feedback loop for quick fixes; (ii) fine-tune to reduce minor syntactic and build issues. \\
    \bottomrule
  \end{tabular}
\end{table*}

In order to gain deeper insights into the erroneous codes generated by our top-performing model (GPT-4.1), we conducted a thorough error analysis. Table \ref{tab:failure} depicts the proportional distribution of each error or failure category. It was observed that the most predominant fault derived from missing or outdated header/module references. API-mismatch and renamed-method issues were commonly observed, often involving deprecated helper functions. Type mismatches, pointer logic errors, and occasional access to private APIs also appeared frequently. Some programs were compiled but failed at runtime, while a small portion exhibited miscellaneous issues such as unused variables, configuration errors, and minor syntax problems. To mitigate these issues in future work, the recommended measures are also described in Table \ref{tab:failure}.

A deeper analysis across difficulty levels and topics reveals important trends. Advanced Wi-Fi tasks contributed the most to the failures, primarily due to missing headers or API mismatches caused by recent changes in Wi-Fi helper interfaces. Tasks involving emerging technologies such as 5G, NR, or mmWave often led to hallucinated headers not part of standard ns-3 builds. In contrast, introductory-level tasks encountered significantly fewer issues, with most errors limited to simple missing \verb|#include|.


\subsection{Limitations \& Future Work}
While SIMCODE offers a strong foundation for evaluating LLMs on network simulation code generation, there are opportunities for further enhancement. The current dataset includes tasks covering a wide range of networking topics; however, it does not yet include examples from emerging domains such as 5G, 6G, IoT, and next-generation wireless systems. Expanding these areas could increase the benchmark’s relevance to cutting-edge simulation tasks. Similarly, while the current evaluation metrics provide a robust measure of code similarity and functional correctness, incorporating additional dimensions such as efficiency, scalability, readability, and alignment with ns-3 best practices would offer a more comprehensive assessment. Finally, although SIMCODE is currently focused on ns-3 simulations, extending its scope to include other simulation platforms (e.g., OMNeT++\footnote{\url{https://omnetpp.org/}}, OPNET\footnote{\url{https://opnetprojects.com/opnet-network-simulator/}}) and programming languages could broaden its applicability and impact.

\section{Conclusion}
\label{sec:conclusion}
This study addressed the critical need for evaluating LLMs in generating domain-specific code for network simulation using ns-3, where the lack of standardized benchmarks has hindered progress in automated code generation. We introduced SIMCODE, comprising tasks across varied topics and difficulty levels, each with prompts, reference solutions, and test cases. Our evaluation across several LLMs showed that GPT-4.1 consistently outperformed others, particularly with CoT prompts. Fine-tuning further improved its performance, demonstrating the value of task-specific adaptation. However, execution accuracy varied widely across topics, with persistent challenges in emerging tech simulations. Error analysis revealed common failure patterns such as outdated headers and hallucinated APIs. These findings highlight the importance of better prompt structuring, feedback-informed, and retrieval-based generation.

Nonetheless, the benchmark currently excludes domains like 6G or IoT and focuses solely on C++, which may limit generalizability. Future work may expand coverage, integrate additional metrics, and explore reinforcement or retrieval-based code correction. In essence, SIMCODE provides a reproducible benchmark that supports LLM development and evaluation for simulation-specific code generation, advancing the way for more reliable use in network systems. Upon acceptance of this manuscript, the SIMCODE dataset and the evaluation and analysis code will be publicly available to support reproducibility and facilitate further research in LLM-driven network simulation. The codes and datasets were not shared during the review process to comply with the double-anonymized submission policy.
\bibliographystyle{IEEEtran}
\bibliography{ref}

\begin{thebibliography}{10}
\providecommand{\url}[1]{#1}
\csname url@samestyle\endcsname
\providecommand{\newblock}{\relax}
\providecommand{\bibinfo}[2]{#2}
\providecommand{\BIBentrySTDinterwordspacing}{\spaceskip=0pt\relax}
\providecommand{\BIBentryALTinterwordstretchfactor}{4}
\providecommand{\BIBentryALTinterwordspacing}{\spaceskip=\fontdimen2\font plus
\BIBentryALTinterwordstretchfactor\fontdimen3\font minus \fontdimen4\font\relax}
\providecommand{\BIBforeignlanguage}[2]{{%
\expandafter\ifx\csname l@#1\endcsname\relax
\typeout{** WARNING: IEEEtran.bst: No hyphenation pattern has been}%
\typeout{** loaded for the language `#1'. Using the pattern for}%
\typeout{** the default language instead.}%
\else
\language=\csname l@#1\endcsname
\fi
#2}}
\providecommand{\BIBdecl}{\relax}
\BIBdecl

\bibitem{chen2021evaluating}
M.~Chen, J.~Tworek, H.~Jun, Q.~Yuan, H.~P. D.~O. Pinto, J.~Kaplan, H.~Edwards, Y.~Burda, N.~Joseph, G.~Brockman \emph{et~al.}, ``Evaluating large language models trained on code,'' \emph{arXiv preprint arXiv:2107.03374}, 2021.

\bibitem{austin2021programsynthesislargelanguage}
\BIBentryALTinterwordspacing
J.~Austin, A.~Odena, M.~Nye, M.~Bosma, H.~Michalewski, D.~Dohan, E.~Jiang, C.~Cai, M.~Terry, Q.~Le, and C.~Sutton, ``Program synthesis with large language models,'' 2021. [Online]. Available: \url{https://arxiv.org/abs/2108.07732}
\BIBentrySTDinterwordspacing

\bibitem{raihan2024mojobench}
N.~Raihan, J.~Santos, and M.~Zampieri, ``Mojobench: Language modeling and benchmarks for mojo,'' \emph{arXiv preprint arXiv:2410.17736}, 2024.

\bibitem{raihan2024mhumaneval}
N.~Raihan, A.~Anastasopoulos, and M.~Zampieri, ``mhumaneval--a multilingual benchmark to evaluate large language models for code generation,'' \emph{arXiv preprint arXiv:2410.15037}, 2024.

\bibitem{riley2010ns}
G.~F. Riley and T.~R. Henderson, ``The ns-3 network simulator,'' in \emph{Modeling and tools for network simulation}.\hskip 1em plus 0.5em minus 0.4em\relax Springer, 2010, pp. 15--34.

\bibitem{genonet2024}
F.~Rezazadeh, A.~A. Gargari, S.~Lagén, J.~Mangues-Bafalluy, D.~Niyato, and L.~Liu, ``Genonet: Generative open xg network simulation with multi-agent llm and ns-3,'' in \emph{2024 3rd International Conference on 6G Networking (6GNet)}, 2024, pp. 69--71.

\bibitem{rezazadeh2025toward}
F.~Rezazadeh, A.~A. Gargari, S.~Lagen, H.~Song, D.~Niyato, and L.~Liu, ``Toward generative 6g simulation: An experimental multi-agent llm and ns-3 integration,'' \emph{arXiv preprint arXiv:2503.13402}, 2025.

\bibitem{manning2022researchagenda}
\BIBentryALTinterwordspacing
S.~Manning, P.~Mishkin, G.~Hadfield, T.~Eloundou, and E.~Eisner, ``A research agenda for assessing the economic impacts of code generation models,'' OpenAI, Research Agenda, March 2022, working paper. [Online]. Available: \url{https://cdn.openai.com/papers/Economic_Impacts_Research_Agenda.pdf}
\BIBentrySTDinterwordspacing

\bibitem{codenet2021}
\BIBentryALTinterwordspacing
R.~Puri, D.~S. Kung, G.~Janssen, W.~Zhang, G.~Domeniconi, V.~Zolotov, J.~Dolby, J.~Chen, M.~R. Choudhury, L.~Decker, V.~Thost, L.~Buratti, S.~Pujar, and U.~Finkler, ``Project codenet: {A} large-scale {AI} for code dataset for learning a diversity of coding tasks,'' \emph{CoRR}, vol. abs/2105.12655, 2021. [Online]. Available: \url{https://arxiv.org/abs/2105.12655}
\BIBentrySTDinterwordspacing

\bibitem{fakhoury2024}
\BIBentryALTinterwordspacing
S.~Fakhoury, A.~Naik, G.~Sakkas, S.~Chakraborty, and S.~K. Lahiri, ``Llm-based test-driven interactive code generation: User study and empirical evaluation,'' \emph{IEEE Transactions on Software Engineering}, vol.~50, no.~9, p. 2254–2268, Sep. 2024. [Online]. Available: \url{http://dx.doi.org/10.1109/TSE.2024.3428972}
\BIBentrySTDinterwordspacing

\bibitem{huang2024agentcodermultiagentbasedcodegeneration}
\BIBentryALTinterwordspacing
D.~Huang, J.~M. Zhang, M.~Luck, Q.~Bu, Y.~Qing, and H.~Cui, ``Agentcoder: Multi-agent-based code generation with iterative testing and optimisation,'' 2024. [Online]. Available: \url{https://arxiv.org/abs/2312.13010}
\BIBentrySTDinterwordspacing

\bibitem{madaan2023selfrefineiterativerefinementselffeedback}
\BIBentryALTinterwordspacing
A.~Madaan, N.~Tandon, P.~Gupta, S.~Hallinan, L.~Gao, S.~Wiegreffe, U.~Alon, N.~Dziri, S.~Prabhumoye, Y.~Yang, S.~Gupta, B.~P. Majumder, K.~Hermann, S.~Welleck, A.~Yazdanbakhsh, and P.~Clark, ``Self-refine: Iterative refinement with self-feedback,'' 2023. [Online]. Available: \url{https://arxiv.org/abs/2303.17651}
\BIBentrySTDinterwordspacing

\bibitem{kumar2025tfhecoderevaluatingllmagenticfully}
\BIBentryALTinterwordspacing
M.~Kumar, J.~Xue, M.~Zheng, and Q.~Lou, ``Tfhe-coder: Evaluating llm-agentic fully homomorphic encryption code generation,'' 2025. [Online]. Available: \url{https://arxiv.org/abs/2503.12217}
\BIBentrySTDinterwordspacing

\bibitem{langchain2024}
\BIBentryALTinterwordspacing
V.~Mavroudis, ``{LangChain v0.3},'' Dec. 2024, working paper or preprint. [Online]. Available: \url{https://hal.science/hal-04817573}
\BIBentrySTDinterwordspacing

\bibitem{hendrycks2021measuring}
\BIBentryALTinterwordspacing
D.~Hendrycks, S.~Basart, S.~Kadavath, M.~Mazeika, A.~Arora, E.~Guo, C.~Burns, S.~Puranik, H.~He, D.~Song, and J.~Steinhardt, ``Measuring coding challenge competence with {APPS},'' in \emph{Thirty-fifth Conference on Neural Information Processing Systems Datasets and Benchmarks Track (Round 2)}, 2021. [Online]. Available: \url{https://openreview.net/forum?id=sD93GOzH3i5}
\BIBentrySTDinterwordspacing

\bibitem{10.1145/3690635}
\BIBentryALTinterwordspacing
J.~Li, G.~Li, Y.~Li, and Z.~Jin, ``Structured chain-of-thought prompting for code generation,'' \emph{ACM Trans. Softw. Eng. Methodol.}, vol.~34, no.~2, Jan. 2025. [Online]. Available: \url{https://doi.org/10.1145/3690635}
\BIBentrySTDinterwordspacing

\bibitem{10.1145/3672456}
\BIBentryALTinterwordspacing
X.~Jiang, Y.~Dong, L.~Wang, Z.~Fang, Q.~Shang, G.~Li, Z.~Jin, and W.~Jiao, ``Self-planning code generation with large language models,'' \emph{ACM Trans. Softw. Eng. Methodol.}, vol.~33, no.~7, Sep. 2024. [Online]. Available: \url{https://doi.org/10.1145/3672456}
\BIBentrySTDinterwordspacing

\bibitem{10.5555/3600270.3602070}
J.~Wei, X.~Wang, D.~Schuurmans, M.~Bosma, B.~Ichter, F.~Xia, E.~H. Chi, Q.~V. Le, and D.~Zhou, ``Chain-of-thought prompting elicits reasoning in large language models,'' in \emph{Proceedings of the 36th International Conference on Neural Information Processing Systems}, ser. NIPS '22.\hskip 1em plus 0.5em minus 0.4em\relax Red Hook, NY, USA: Curran Associates Inc., 2022.

\bibitem{yao2023reactsynergizingreasoningacting}
\BIBentryALTinterwordspacing
S.~Yao, J.~Zhao, D.~Yu, N.~Du, I.~Shafran, K.~Narasimhan, and Y.~Cao, ``React: Synergizing reasoning and acting in language models,'' 2023. [Online]. Available: \url{https://arxiv.org/abs/2210.03629}
\BIBentrySTDinterwordspacing

\bibitem{wang2023selfconsistencyimproveschainthought}
\BIBentryALTinterwordspacing
X.~Wang, J.~Wei, D.~Schuurmans, Q.~Le, E.~Chi, S.~Narang, A.~Chowdhery, and D.~Zhou, ``Self-consistency improves chain of thought reasoning in language models,'' 2023. [Online]. Available: \url{https://arxiv.org/abs/2203.11171}
\BIBentrySTDinterwordspacing

\bibitem{ren2020codebleumethodautomaticevaluation}
\BIBentryALTinterwordspacing
S.~Ren, D.~Guo, S.~Lu, L.~Zhou, S.~Liu, D.~Tang, N.~Sundaresan, M.~Zhou, A.~Blanco, and S.~Ma, ``Codebleu: a method for automatic evaluation of code synthesis,'' 2020. [Online]. Available: \url{https://arxiv.org/abs/2009.10297}
\BIBentrySTDinterwordspacing

\end{thebibliography}

\end{document}